\documentclass[preprint]{aastex62}            

\usepackage{amsmath}
\usepackage{bm}
\usepackage{multirow}
\usepackage{xcolor}

\begin{document}

\title{Photodissociation of CS from Excited Rovibrational Levels}

\author{R. J. Pattillo}
\affil{Department of Physics and Astronomy and Center for Simulational Physics \\ 
        The University of Georgia, Athens, GA 30602-2451, USA}
\email{ryanpattillo@uga.edu}

\author{R. Cieszewski}
\altaffiliation{Deceased}
\affil{Department of Physics and Astronomy and Center for Simulational Physics \\ 
        The University of Georgia, Athens, GA 30602-2451, USA}

\author{P. C. Stancil}
\affil{Department of Physics and Astronomy and Center for Simulational Physics \\ 
        The University of Georgia, Athens, GA 30602-2451, USA}
\email{stancil@physast.uga.edu}

\author{R. C. Forrey}
\affil{Department of Physics, Penn State University, Berks Campus, Reading, PA 19610, USA}
\email{rcf6@psu.edu}

\author{J. F. Babb}
\affil{ITAMP, Harvard-Smithsonian Center for Astrophysics \\
MS 14, 60 Garden Street, Cambridge, MA 02138-1516, USA}
\email{jbabb@cfa.harvard.edu}

\author{J. F. McCann}
\affil{Centre for Theoretical Atomic, Molecular, and Optical Physics, School of Mathematics \& Physics \\ 
Queen's University of Belfast, Belfast BT7 1NN, UK}
\email{j.f.mccann@icloud.com}

\author{B. M. McLaughlin}
\affil{Centre for Theoretical Atomic, Molecular, and Optical Physics, School of Mathematics \& Physics \\ 
Queen's University of Belfast, Belfast BT7 1NN, Northern Ireland, UK}
\email{bmclaughlin899@btinternet.com}

\begin{abstract}
Accurate photodissociation cross sections have been computed for transitions 
from the $\text{X}\ ^1\Sigma^+$ ground electronic state of CS to six low-lying excited
electronic states. New \textit{ab initio} potential curves and transition dipole moment functions 
have been obtained for these computations using the multi-reference configuration interaction 
approach with the Davidson correction (MRCI+Q) and aug-cc-pV6Z basis sets. State-resolved 
cross sections have been computed for transitions from nearly the full range of rovibrational 
levels of the $\text{X}\ ^1\Sigma^+$ state and for photon wavelengths ranging from 500 \AA\ to 
threshold. {  Destruction of CS via predissociation in highly excited electronic states originating from the rovibrational ground state is found to be unimportant.} Photodissociation cross sections are presented for temperatures in the range between 
1000 K and 10,000 K, where a Boltzmann distribution of initial rovibrational levels is assumed. 
Applications of the current computations to various astrophysical environments are briefly
 discussed focusing on photodissociation rates 
due to the standard interstellar and blackbody radiation fields.
\end{abstract}

\keywords{astrochemistry --- molecular data --- molecular processes}

\section{Introduction}\label{sec:introduction}

{ CS is a molecule of great astrophysical interest.
It is also one of the most abundant  sulfur-bearing compounds in interstellar clouds and}
is found in a variety of astrophysical objects including star-forming 
regions \citep{walker86}, protostellar envelopes \citep{her12}, 
dense interstellar clouds \citep{hasegawa84,hayashi85,des09},
 carbon-rich stars \citep{bregman78,ridgway77,tenenbaum10}, 
oxygen-rich stars \citep{ziurys07,tenenbaum10}, planetary nebulae \citep{edwards14}, 
and comets \citep{smith80,jackson82,canaves07}. 

Photodissociation is an important mechanism for the destruction of molecules in
environments with an intense radiation field, so accurate photodissociation rates 
are necessary to estimate the abundance of CS. \citet{heays17} 
presented photodissociation cross 
sections and photorates for CS using previous  estimates \citep{vandishoeck88}
{ applying}
 measured wavelengths for transitions to the B$~^1\Sigma^+$ (or 3$~^1\Sigma^+$)
 from the ground state \citep{stark87} and vertical excitation energies of higher states \citep{bruna75}.
However, 
comprehensive photodissociation cross sections are needed 
to compute photorates in many environments. In response, we have calculated 
{ photodissociation}
cross sections for 
the CS molecule for several electronic transitions from a wide range of initial rovibrational levels. 
Photodissociation cross sections for transitions from the $\text{X}\ ^1\Sigma^+$ electronic 
ground state to the $\text{A}\ ^1\Pi$, $\text{A}^{\prime}\ ^1\Sigma^+$($2\ ^1\Sigma^+$), 
$2\ ^1\Pi$, $3\ ^1\Pi$, $\text{B}\ ^1\Sigma^+$($3\ ^1\Sigma^+$), 
and $4\ ^1\Pi$ electronic states are studied here. Calculations have been performed for transitions 
from $14,908$ initial bound rovibrational levels $v'',J''$ of the X state. { We also explore predissociation out of the $3\ ^1\Pi$, $\text{B}\ ^1\Sigma^+$, 
and $4\ ^1\Pi$ excited electronic states.}

The present cross section calculations 
are
performed using quantum-mechanical techniques. 
Applications of the cross sections to environments appropriate for local thermodynamic equilibrium (LTE) 
conditions are included, where a Boltzmann distribution of initial rovibrational levels is assumed.
Photodissociation rates 
are
computed for the standard interstellar radiation field (ISRF) 
and for blackbody radiation fields at a wide range of temperatures. 

The layout of this paper is as follows. An overview of the theory of molecular photodissociation 
and the adopted molecular data is presented in \autoref{sec:theory}. In \autoref{sec:results}, 
the computed state-resolved cross sections, LTE cross sections, and photodissociation rates are 
discussed. Finally, in \autoref{sec:conclusions}, conclusions are drawn from our work. 
Atomic units are used throughout unless otherwise specified.

\section{Theory and Calculations}\label{sec:theory}

\subsection{Potential Curves and Transition Dipole Moments}\label{sec:pecstdms}

In a similar manner to our recent molecular structure work on the SiO molecule \citep{forrey16,forrey17},
 which is iso-electronic to CS, the potential energy curves and transition dipole matrix (TDM) elements for 
several of the low-lying electronic states are calculated. We use a 
state-averaged-multi-configuration-self-consistent-field (SA-MCSCF) approach, followed by multi-reference
configuration interaction (MRCI) calculations together with the Davidson correction \citep[MRCI+Q;][]{Helgaker2000}. 
The SA-MCSCF method is used as the reference wave function for the MRCI calculations. 
  
Potential energy curves (PECs) and TDMs as a function of internuclear distance $R$ are calculated starting from a 
bond separation of $R = 1.8$ Bohr extending out to $R=12$ Bohr. At bond distances beyond this value we use a multipole expansion, 
detailed below, to represent the long-range part of the potentials. The basis sets used in our work are the 
augmented correlation consistent polarized sextuplet (aug-cc-pV6Z (AV6Z)) Gaussian basis sets. 
The use of such large basis sets is
well known to recover 98\% of 
the electron correlation effects in molecular structure calculations \citep{Helgaker2000}. 
All the PEC and TDM calculations for the CS molecule were performed with the quantum chemistry 
program package MOLPRO 2015.1 \citep{molpro2015}, running on parallel architectures.
	
{
\floattable
\begin{deluxetable}{l l C C C}[ht!]
 \tablewidth{0pt}
    \tablecaption{The permanent dipole moment $\mu_X$ for the  $\text{X}\ ^1\Sigma^+$ ground state of the 
                           CS molecule at $2.9$ $a_0$/1.5346 \AA, a value near equilibrium, compared with experiment, 
                           SCF, CAS-CI, MRCI+Q, and MCSCF theoretical calculations.\label{table:pdm}}
  \tablehead{ \text{CS Ground State}& \text{Method} & \text{Basis Set} & \mu_X\ \text{(Debye)} & \Delta\ (\%) }
  \startdata
 $\text{X}\ ^1\Sigma^+$  	&\text{EXPT$^{a}$}		& \nodata 	 								&1.958 \pm 0.005& \nodata \\
 \nodata  				&\text{MRCI+Q$^{b}$} 	 	& \text{aug-cc-pV6Z} 						&2.042 			& +{4.3}\% \\
 \nodata   				&\text{MCSCF$^{c}$}			& \text{aug-cc-pV6Z}						&2.179 			& +11\% \\
 \nodata   				&\text{CAS-CI$^{d}$} 	 	& \text{double-zeta + polarization (DZP)}	&2.147 			& +{9.7}\% \\
 \nodata 				& \text{SCF$^{e}$}			& \text{double-zeta + polarization (DZP)} 	&1.783 			& -{8.9}\% \\
 \nodata 				& \text{CI$^{f}$}			& \text{double-zeta (DZ)}		 			&2.350 			& +{20}\% \\
 \nodata 				& \text{HF$^{g}$}			& \text{double-zeta (DZ)} 					&1.650 			& {-16}\% \\
\enddata
 \tablenotetext{a}{Experiment \citep{winnewisser68}.}
 \tablenotetext{b}{Multi-reference configuration interaction with the Davidson correction (MRCI+Q; present work).}
 \tablenotetext{c}{Multi-configuration-self-consistent-field (MCSCF; present work).}
 \tablenotetext{d}{Complete-active-space configuration interaction (CAS-CI), with the SWEDEN codes (present work).}
 \tablenotetext{e}{Self-consistent field \citep[SCF;][]{varambhia10}.}
 \tablenotetext{f}{Configuration interaction \citep[CI;][]{rs76}.}
 \tablenotetext{g}{Hartree-Fock \citep[HF;][]{rs76}.}
\end{deluxetable}
}
	
For molecules with degenerate symmetry, an Abelian subgroup is required to be used
{ in MOLPRO}. 
For a diatomic 
molecule like CS with C$_{{\infty}v}$ symmetry, it will be substituted by C$_{2v}$ symmetry with the 
order of irreducible representations being ($A_1$, $B_1$, $B_2$,  $A_2$). When symmetry is reduced 
from C$_{{\infty}v}$ to C$_{2v}$, the correlating relationships are 
$\sigma \rightarrow a_1$, $\pi \rightarrow$ ($b_1$, $b_2$), and $\delta \rightarrow$ ($a_1$, $a_2$). 
In order to take account of short-range interactions, we employed the non-relativistic state-averaged 
complete active-space-self-consistent-field (SA-CASSCF)/MRCI method 
available within the MOLPRO \citep{molpro2012,molpro2015} quantum chemistry suite of codes.  

{
\floattable
\begin{deluxetable}{llccc}[ht!]
    \tablewidth{0pt}
\tabletypesize{\scriptsize}
\tablecaption{Equilibrium Bond distance $R_e$ (\AA{}) and
             Dissociation Energies $D_e$ (eV) for the  $\text{X}\ ^1\Sigma^+$, 
			$\text{A$^{\prime}$}\ ^1\Sigma^+$ and  $\text{A}\ ^1\Pi$	 
  			 Molecular States of CS from
             the Present \textsc{MRCI+Q} Calculations Compared to Other 
             Theoretical and Experimental Results
         \label{table:constants}}
    \tablehead{Molecular State 	& Method  				& Basis Set							& $R_e$/\AA{} 	& $D_e$/(eV)}
 \startdata
 $\text{X}\ ^1\Sigma^+$ 		&MRCI + Q$^a$			& aug-cc-pV6Z (AV6Z) 				& 1.5314 	 	& 7.6113 \\
								&MRCI + Q$^b$			& aug-cc-pwCV5Z (ACV5Z)				& 1.5346		& 7.3851 \\
								&MRCI + Q$^c$			& aug-cc-pV6Z (AV6Z)				& 1.5346		& \nodata \\
								&MRCI + Q + cv + dk$^d$	& aug-cc-pV6Z (AV6Z)				& 1.5377		& \nodata \\
								&CCSD(T) + cv + dk + 56$^e$	& aug-cc-pV6Z (AV6Z)			& 1.5387		& \nodata \\
                                &MRCI$^f$				& aug-cc-pC5VZ (C) + aug-cc-pV5Z (S)&1.5334 		& 7.3436\\
                             	&M-S-APEF$^g$   		& aug-cc-pC5VZ (C) + aug-cc-pV5Z (S)&1.5403 		& 7.3436\\
                                &HF/DF-B3LYP$^h$		& aug-cc-pVTZ						& 1.5360		& 7.0644\\
								&EXPT$^i$				& \nodata							& 1.5349       	& \nodata\\
								&EXPT$^j$				& \nodata 							& 1.5350       	& \nodata\\
								&EXPT$^k$				& \nodata                  			& 1.5349         & 7.3530 $\pm$ 0.025\\
								&MORSE/RKR$^l$			& \nodata							& 1.5349		& 7.4391\\
\\
 A$^{\prime}\ ^1\Sigma^+(\text{2}\ ^1\Sigma^+$) &MRCI + Q$^a$& aug-cc-pV6Z (AV6Z) 			& 1.9443 	 	&0.4558 \\
								&MRCI + Q$^b$			& aug-cc-pwCV5Z (ACV5Z)				& 1.9399        &0.4253 \\
								&MRCI + Q$^c$			& aug-cc-pV6Z (AV6Z)				& 1.9399       	& \nodata\\
								&EXPT$^i$				& \nodata 							& 1.9440       	& \nodata\\
								&EXPT$^j$				& \nodata 							& 1.9440 	   	& \nodata\\
\\
 $\text{A}\ ^1\Pi$	 			&MRCI + Q$^a$			& aug-cc-pV6Z (AV6Z) 				& 1.5622 	 	& 2.7333\\
								&MRCI + Q$^b$			& aug-cc-pwCV5Z (ACV5Z)				& 1.5676		& 2.6637\\						
								&MRCI + Q$^c$			& aug-cc-pV6Z (AV6Z)				& 1.5676		& \nodata\\	
								&MRCI + Q + cv + dk$^d$	& aug-cc-pV6Z (AV6Z)				& 1.5690		& \nodata\\	
								&EXPT$^i$				& \nodata 							& 1.5739       	& \nodata\\
								&EXPT$^j$				& \nodata							& 1.5660       	& \nodata\\
\enddata
{\tiny  \tablecomments{The data are given in units conventional to quantum chemistry with $1~\text{\AA} = 10^{-10}~\text{m}$ and $0.529177~\text{\AA} \approx 1~a_0$.
	The conversion factor $1.239842 \times 10^{-4}~\text{eV} = 1~\text{cm}^{-1}$ is also used.}
	\tablenotetext{a}{MRCI+Q, Multi-reference configuration interaction (MRCI) with Davidson correction (Q; present work).}
  \tablenotetext{b}{MRCI+Q, ACV5Z  \citep{li13}}
  \tablenotetext{c}{MRCI+Q, AV6Z \citep{shi11}}
  \tablenotetext{d}{MRCI+Q+cv+dk, core-valence (cv) and relativistic effects \citep[dk;][]{shi11}}
  \tablenotetext{e}{CCSD(T)+cv+dk+56, Coupled cluster (CCSD(T)), core-valence, 
                    relativistic effects/basis set limit \citep{shi11}}
  \tablenotetext{f}{MRCI \citep{shi10}}
  \tablenotetext{g}{M-S-APEF, Murrell-Sobbell (M-S) fit with 
  analytic potential energy function \citep[APEF;][]{shi10}}
  \tablenotetext{h}{HF/DF-B3LYP, Hybrid density functional method \citep{das03}}
  \tablenotetext{i}{Experiment \citep{Huber1979}}
  \tablenotetext{j}{Experiment \citep{bergeman81}}
  \tablenotetext{k}{Experiment \citep{coppens95}}
  \tablenotetext{l}{Morse with Rydberg-Klein-Rees (RKR) potential \citep{nadhem15}}
}
\vspace{-10pt}
\end{deluxetable}
}

For the CS molecule, eight molecular orbitals (MOs) are put into the active space, including 
four $a_1$, two $b_1$ and two $ b_2$ symmetry MOs which correspond to the $3s3p$ shell of 
sulfur and $2s2p$ shell of carbon. The rest of the electrons in the CS molecule 
are put into closed-shell orbitals, including four $a_1$, one $b_1$ and one $b_2$ symmetry MOs. 
The molecular orbitals for the MRCI procedure were obtained using the SA-MCSF method, for which we
 carried out the averaging processes on the lowest three $^1\Sigma^+$ ($^1A_1$), three $^1\Pi$ ($^1B_1$), 
three $^3\Sigma^+$ ($^3A_1$), three  $^3\Pi$ ($^3B_1$), two $^1\Delta$ ($^1A_2$) and 
two $^3\Delta$ ($^3A_2$) molecular states. The fourteen MOs ($8a_1$, 3b$_1$, 3b$_2$, $0a_2$), i.e. (8,3,3,0), 
were then used to perform all the PEC and TDM calculations for the electronic states of interest in the 
MRCI+Q approximation. \autoref{table:pdm} compares theoretical results for the permanent dipole moment 
$\mu_X$ of the $\text{X}\ ^1\Sigma^+$ ground state at various levels of approximation with experiment 
to demonstrate the accuracy of the MRCI+Q approximation applied here.  
As can be seen from  \autoref{table:pdm} our MRCI+Q 
results for the permanent dipole moment of CS, for the ground state, at the equilibrium geometry, are within 4\% of 
the experimental value \citep{winnewisser68}. \autoref{table:constants} compares 
the equilibrium distance $R_e$ (\AA{}) and the dissociation energy $D_e$ (eV) 
at various level of approximation for the $\text{X}~^1\Sigma^+$, $\text{A}^{\prime}\ ^1\Sigma^+$,  and A$~^1\Pi$ states of CS.   
We note that for the  $\text{X}\ ^1\Sigma^+$ ground state, 
the early experimental work of \cite{crawford34} determined values 
 $R_e$ = 1.2851 \AA{} and $D_e$ (eV) = 7.752 eV, 
in less favorable agreement 
with our present {\it ab initio} work or that of other high level molecular structure calculations. 
As shown in \autoref{table:constants} the use of polarized-core-valence basis 
sets by \citet{li13} provides spectroscopically accurate results 
for $R_e$ (\AA{}) and $D_e$ (eV),
respectively, being within 0.03 pm and 0.032 eV 
compared with the available experiment.
{ We find that our TDMs differ slightly in magnitude but agree in trend with those presented by \citet{li13} on the range they are computed.}

Beyond a bond separation of $R = 12$ Bohr, a multipole expansion is smoothly fitted to the PECs and TDMs up to $R = 100$ Bohr.
For the PECs this has the form
\begin{equation}\label{eq:longrange}
    V(R) = -\frac{C_5}{R^5}-\frac{C_6}{R^6}\,,
\end{equation}
where $C_5$ and $C_6$ are 
coefficients for each electronic state shown in \autoref{table:states}. 
For $R < R_{\textrm{min}}$, down to a bond length of $1.5\ \text{a}_0$, a short-range interaction potential 
of the form $V(R) = A \exp(-BR)+C$ was fitted to the \textit{ab initio} potential curves. 

A method to estimate the value of the quadrupole-quadrupole
coefficient $C_5$  for an electronic state of a diatomic molecule 
like CS is given by \citet{chang67}. In order to compute the long-range dispersion coefficient $C_6$, the London formula
\begin{equation}
    C_6 = \frac{3}{2}\frac{\cal{I_{\text{C}}I_{\text{S}}}}{\cal{[I_{\text{C}}+ I_{\text{S}}]}}\alpha_{\text{C}}\alpha_{\text{S}}
\end{equation}
is applied, where $\alpha$ is the dipole polarizability and $\cal{I}$ is the ionization energy of each of the atoms 
in a given atomic state. The ionization energies are taken from the NIST Atomic Spectra Database \citep{NIST}. 
For the sulfur atom, the dipole polarizabilities of $\alpha_{\text{S}} = 18.8$ and $\alpha_{\text{S}} = 19.5$,
respectively,
are used for the ground state $3s^23p^4\ ^3P$ and the excited state $3s^23p^4\ ^1D$ 
\citep{mukherjee89}. 
For the ground state $2s^22p^2\ ^3P$ of atomic carbon, a dipole polarizability of $\alpha_{\text{C}} = 10.39$ is 
used \citep{miller72}. An estimated value of $\alpha_{\text{C}} = 10.78$ is used for the excited state $2s^22p^2\ ^1D$ 
of carbon: this value was obtained by scaling the ground state polarizability 
to match the ratio of the $^3P$ and $^1D$ polarizabilities of sulfur. 

{
\floattable
\begin{deluxetable}{c@{$~$}L L@{\ +\ }L D D D D L}[ht!]
    \tablewidth{0pt}
    \tablecaption{CS Electronic States\label{table:states}}
    \tablehead{\multicolumn{2}{c}{Molecular} & \multicolumn{9}{c}{Separated Atom}  & &\multicolumn{1}{c}{United Atom}\\
    \cline{3-11}
    \multicolumn{2}{c}{State} & \multicolumn{2}{c}{Atomic State} &
    \multicolumn{2}{c}{Energy (eV)\tablenotemark{a}} &
    \multicolumn{2}{c}{Energy (eV)\tablenotemark{b}} &
    \multicolumn{2}{c}{$C_5$\tablenotemark{c}} &
    \multicolumn{2}{c}{$C_6$\tablenotemark{d}} &
    \multicolumn{1}{c}{State}}
    \decimals
    \startdata
    X&$^1\Sigma^+$ 				& $\text{C}(2s^22p^2\ ^3P) & \text{S}(3s^23p^4\ ^3P)$ & 0.0 	& 0.0 			& 27.34	 	& 58.02 	& $3d^24s^2\ a~^1D$  \\
    A&$^1\Pi$ 					& $\text{C}(2s^22p^2\ ^3P) & \text{S}(3s^23p^4\ ^3P)$ & 0.0 	& 7.95(-4) 		& 0.0 		& 58.02 	& $3d^24s^2\ a~^1D$  \\
    A$^{\prime}$&$^1\Sigma^+$ 	& $\text{C}(2s^22p^2\ ^3P) & \text{S}(3s^23p^4\ ^3P)$ & 0.0 	& 1.14(-3)		& 0.0 		& 58.02 	& $3d^24s^2\ a~^1G$  \\
    2&$^1\Pi$ 					& $\text{C}(2s^22p^2\ ^3P) & \text{S}(3s^23p^4\ ^3P)$ & 0.0 	& 6.52(-4) 		& -18.23 	& 58.02 	& $3d^24s^2\ a~^1G$  \\
    \hline
    B&$^1\Sigma^+$ 				& $\text{C}(2s^22p^2\ ^1D) & \text{S}(3s^23p^4\ ^1D)$ & 2.3812287  & 2.38172  	& 27.26 	& 55.56  	& $3d^34s\ b~^1G$  \\
    3&$^1\Pi$ 					& $\text{C}(2s^22p^2\ ^1D) & \text{S}(3s^23p^4\ ^1D)$ & 2.3812287  & 2.38652 	&10.06 		& 55.56 	& $3d^34s\ b~^1G$ \\
    4&$^1\Pi$ 					& $\text{C}(2s^22p^2\ ^1D) & \text{S}(3s^23p^4\ ^1D)$ & 2.3812287  & 2.39969 	& -11.81 	& 55.56  	& $3d^34s\ a~^1H$  \\
    \enddata
    \tablenotetext{a}{Experimental data from NIST Atomic Spectra Database \citep{NIST}.}
    \tablenotetext{b}{Current theory extrapolated to the asymptotic limit with Eq.~(\ref{eq:longrange}).} 
    \tablenotetext{c}{Estimated following \citet{chang67}. See the text for details.}
    \tablenotetext{d}{Estimated from the London formula. See the text for details.}
\end{deluxetable}
}

\begin{figure}
    \centering
    \includegraphics[width=\textwidth]{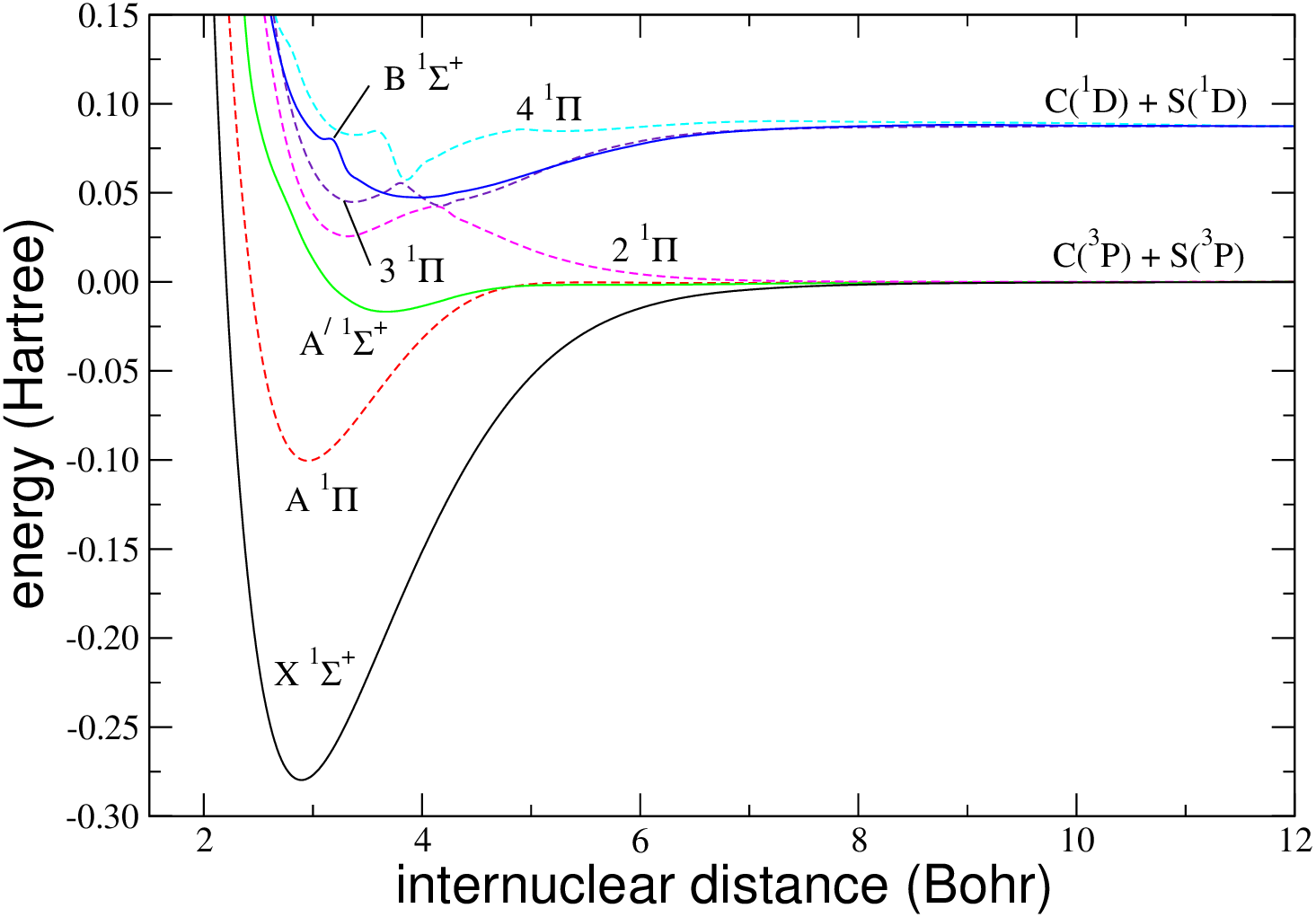}
    \caption{Potential energy curves for each considered CS molecular state.}
    \label{fig:pots}
\end{figure}

The potentials for the excited states of the CS molecule were shifted so that the asymptotic energies as 
$R\rightarrow\infty$ agree with the separated atom energy differences found in the 
NIST Atomic Spectra Database \citep{NIST} shown in \autoref{table:states}. 
Except for the 4~$^1\Pi$ state,
shifts are less than $\sim$5 meV indicating the reliability of the
MRCI+Q calculations within the uncertainty of the estimated dispersion coefficients.
The potential curves for CS 
are shown in \autoref{fig:pots}.

\begin{figure}
    \centering
    \includegraphics[width=\textwidth]{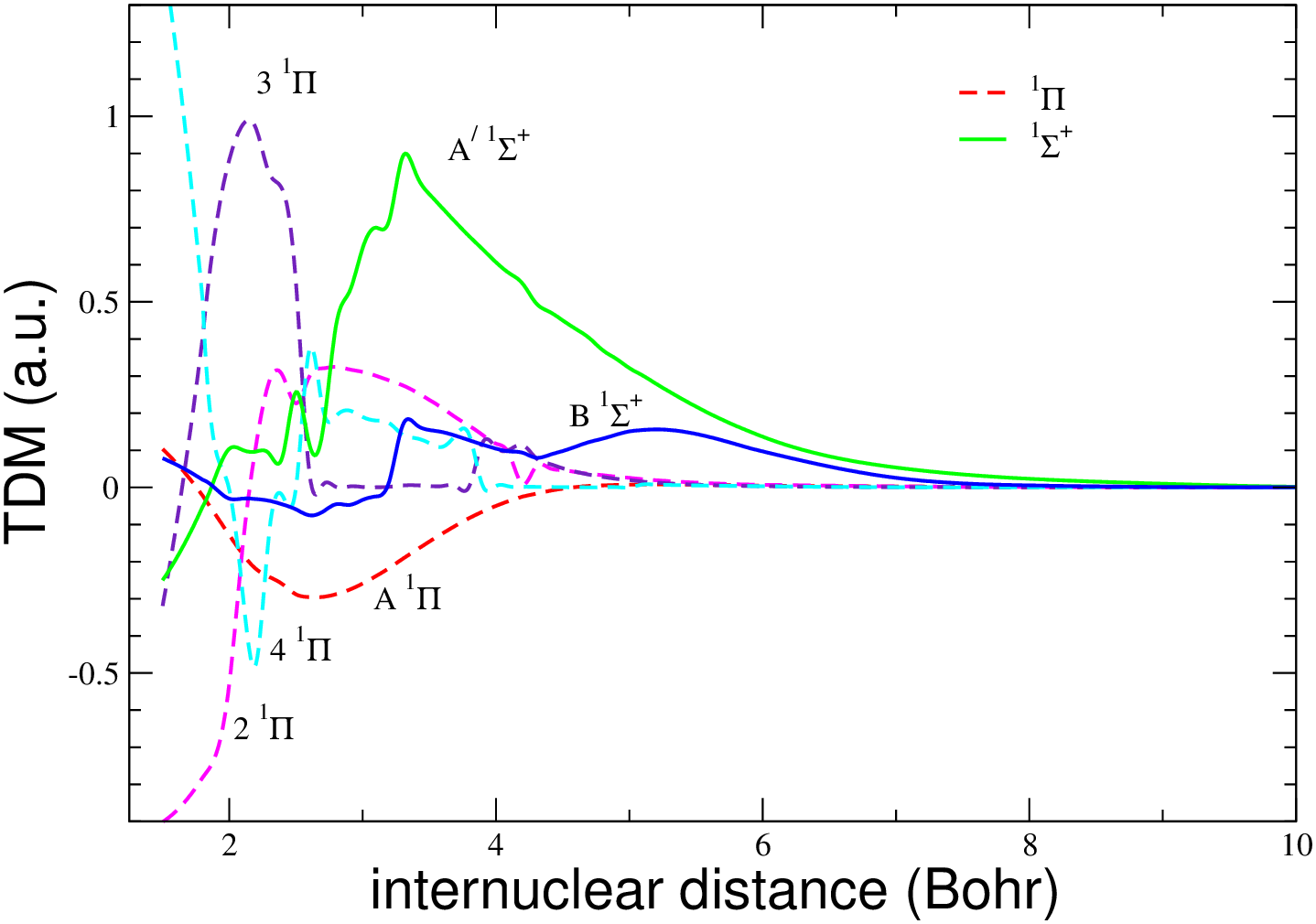}
    \caption{Transition dipole moments for transitions from the ground state to each CS excited state.}
    \label{fig:tdms}
\end{figure}

The TDMs for the CS molecule are similarly extended to long and short-range internuclear bond distances. 
For $R > R_{\textrm{max}}$ a functional fit of the form $D(R) = a \exp(-bR) + c$ is applied, 
while in the short-range $R < R_{\textrm{min}}$ a quadratic fit of the form $D(R) = a' R^2 + b' R + c'$ is adopted. 
We deduce from the atomic states of C and S that the long-range $R \to \infty$ limit of each TDM is zero. 
Similarly, the united-atom limit (which is the Ti atom) as $R \to 0$ of each TDM is zero as well
(see Table~\ref{table:states}). The TDMs are shown in \autoref{fig:tdms}.

The wave functions of the bound rovibrational levels are computed by solving the radial Schr\"{o}dinger 
equation for nuclear motion on the $\text{X}\ ^1\Sigma^+$ potential curve. The wave functions 
are obtained numerically using the standard Numerov method \citep{cooley61,johnson77} with a 
step size of $0.001$ Bohr. We find 85 vibrational levels with a total of 14,908 rovibrational levels. 
This covers nearly the full range of rovibrational levels in the $\text{X}\ ^1\Sigma^+$ state.

\subsection{The Photodissociation Cross Section}

{Here, we present a brief overview of the state-resolved photodissociation cross section calculation; further details are given in previous work \citep{miyake11}. }
In units of cm$^2$,
 the state-resolved cross section for a bound-free transition from initial rovibrational level $v''J''$ is
\begin{equation}
    \sigma_{v''J''} (E_{ph}) = 2.689\times10^{-18} E_{ph} g \sum_{J'} \left( \frac{1}{2J''+1} S_{J'} |D_{k'J',v''J''}|^2 \right) 
\end{equation}
\citep{kirby88} where $k'J'$ are the continuum states of the final electronic state. 
The H\"{o}nl-London factors, $S_{J'}(J'')$ \citep{watson08}, are expressed for a $\Sigma\leftarrow\Sigma$ electronic transition as
\begin{equation}
    S_{J'}(J'') = 
    \begin{cases}
    J'', & J' = J''-1 \phn\phn (\textrm{P-branch}) \\
    J''+1, & J' = J''+1 \phn\phn (\textrm{R-branch})\,,
    \end{cases}
\end{equation}
and for a $\Pi\leftarrow\Sigma$ transition as
\begin{equation}
    S_{J'}(J'') = 
    \begin{cases}
    (J''-1)/2, & J' = J''-1 \phn\phn (\textrm{P-branch}) \\
    (2J''+1)/2, & J' = J'' \phn\phn (\textrm{Q-branch}) \\
    (J''+2)/2, & J' = J''+1 \phn\phn (\textrm{R-branch})\,.
    \end{cases}
\end{equation}
The matrix element of the electric TDM for absorption from $v''J''$ to the continuum $k'J'$ is
\begin{equation}
    D_{k'J',v''J''} = \langle \chi_{k'J'}(R) | D(R) | \chi_{v''J''}(R)  \rangle\,,
\end{equation}
with the integration taken over $R$ where $D(R)$ is the appropriate TDM function. 
The bound rovibrational wave functions $\chi_{v''J''}$ { and continuum wave functions $\chi_{k'J'}(R)$ are} computed using the standard Numerov method 
with a step size of $0.001$ Bohr. They are normalized such that they behave asymptotically as
\begin{equation}
    \chi_{k'J'}(R) \sim \sin \left( k'R - \frac{\pi}{2} J' + \eta_{J'} \right)\ ,
\end{equation}
where $\eta_{J^\prime}$ is the single-channel phase shift of the upper electronic state.
Finally, the degeneracy factor $g$ is given by
\begin{equation}
    g = \frac{2-\delta_{0,\Lambda'+\Lambda''}}{2-\delta_{0,\Lambda''}}\,,
\end{equation}
where $\Lambda'$ and $\Lambda''$ are the angular momenta projected along the
 nuclear axis for the final and initial electronic states, respectively. 

Predissociation is also possible through an intermediate transition to a bound level of an excited state.
In units of cm$^2$, the predissociation cross section is
\begin{equation}
	\sigma = 8.85 \times 10^{-22} \lambda^2 x_{\ell} f_{u\ell} \eta^d
    \label{pred}
\end{equation}
\citep{heays17}, where $\lambda$ is the photon wavelength in \AA\ and $f_{u\ell}$ is the oscillator strength of the transition from lower state $\ell$ to upper state $u$.
We approximate the ground-state fractional population $x_{\ell}$ and the upper level tunneling probability $\eta^d$ to both be 1
to give an upper limit to the predissociation cross section.

\subsection{LTE Cross Sections}

In LTE, a Boltzmann population distribution is assumed for the rovibrational 
levels in the electronic ground state. The total quantum-mechanical photodissociation 
cross section as a function of both temperature $T$ and wavelength $\lambda$ is
\begin{equation}
    \sigma(\lambda,T) = \frac{\sum_{v''}\sum_{J''} g_{iv''J''} \exp [-(E_{00} - E_{v''J''}) / k_b T] \sigma_{v''J''}}{\sum_{v''} \sum_{J''} g_{iv''J''} \exp [ -(E_{00} - E_{v''J''} ) / k_b T]}\,,
\end{equation}
where $g_{iv''J''} = 2J''+1$ is the otal statistical weight, $E_{v''J''}$ is the magnitude of 
the binding energy of the rovibrational level $v''J''$, and $k_b$ is the Boltzmann constant. 
The denominator is the rovibrational partition function.

\subsection{Photodissociation Rates}

The photodissociation rate for a molecule in an ultraviolet radiation field is given by
\begin{equation}
    k = \int \sigma(\lambda)I(\lambda)\,d\lambda\,,
    \label{rate}
\end{equation}
where $\sigma(\lambda)$ is the photodissociation cross section and $I(\lambda)$ is the 
photon radiation intensity summed over all incident angles. The photon radiation 
intensity emitted by a blackbody with temperature $T$ is
	%
	%
    	%
\begin{equation}
    I(\lambda,T) = \frac{8\pi c/\lambda^4}{\exp(hc/k_b T \lambda) - 1} 
    \label{planckrad}
\end{equation}
where $h$ is the Planck constant and $c$ is the speed of light. 

 We also compute the photodissociation rate in the unattenuated ISRF, as given by \citet{draine}, but
modified for $\lambda > 2000$ \AA\ by \citet{heays17}, using Eq.~(\ref{rate}). In an interstellar cloud the radiation
field is attenuated by dust reducing the photodissociation rate as a function of depth into the cloud, or parameterized
as the visual extinction $A_{\rm V}$. Assuming a plane-parallel, semi-infinite slab, with { both sides of the cloud
exposed isotropically} to the ISRF, we applied the radiative transfer code of \citet{roberge} to compute the photodissociation
rate as a function of $A_{\rm V}$ and fit the rate to the forms
\begin{equation}
k(A_{\rm V}) = a_1\exp{(-a_2A_{\rm V} +a_3 A_{\rm V}^2)},
\label{fit1}
\end{equation}
{
\begin{equation}
k(A_{\rm V}) = a_4E_2(a_5A_{\rm V}),
\label{fit2}
\end{equation}
where $E_2$ is the second-order exponential integral.}
The grain model of \citet{dralee} that was adopted corresponds to the galactic average of the total-to-selective
extinction $R_{\rm V} = 3.1$.

\section{Results and Discussion}\label{sec:results}

\subsection{State-resolved Cross Sections}

\begin{figure}
    \centering
     \includegraphics[width=\textwidth]{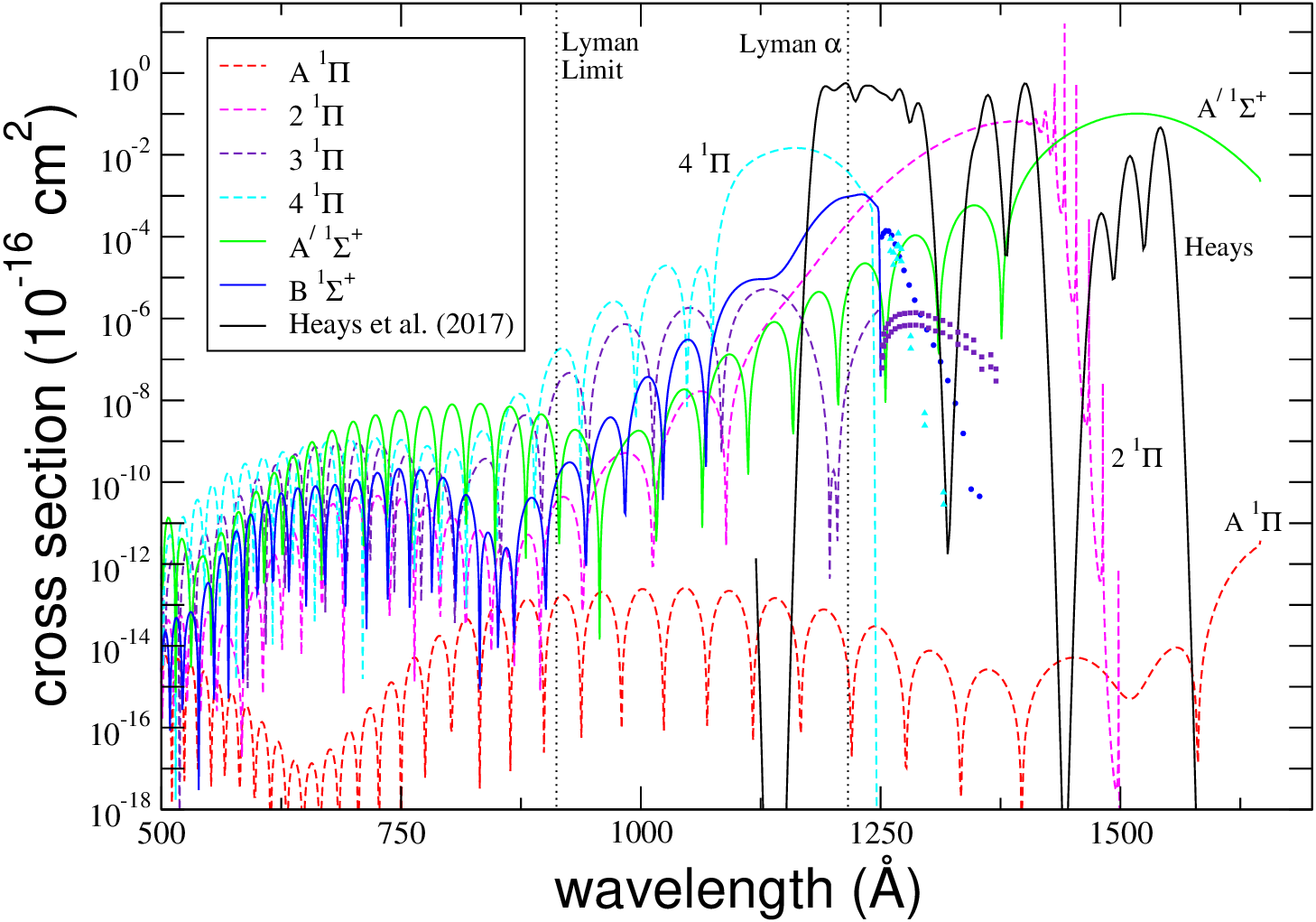}
  \caption{Comparison of CS state-resolved cross sections for transitions from the ground rovibrational level $v'',J'' = 0,0$.
	The CS cross section estimate of \citet{vandishoeck88}, 
as adopted in \citet{heays17}, is shown for comparison. { Continuum cross sections (solid lines); predissociation (points) longward of $\sim$1250 \AA.}}
    \label{fig:v0j0}
\end{figure}

State-resolved photodissociation cross sections have been computed for transitions from 14,908 initial 
rovibrational levels in the $\text{X}\ ^1\Sigma^+$ ground electronic state to the six considered 
excited electronic states. Cross sections are computed for photons with wavelengths starting at 
500 \AA\ up to at most 50,000 \AA\ in 1 \AA\ increments, typically stopping at the relevant threshold. { A smaller wavelength step size is used near thresholds to resolve appropriate resonances.} 
In \autoref{fig:v0j0}, a comparison of the state-resolved cross sections from the ground rovibrational 
level $v'',J''=0,0$ for each transition is shown. 
The $2\ ^1\Pi$ and  $\text{A}^{\prime}\ ^1\Sigma^+$ ($2\ ^1\Sigma^+$) transitions have 
the dominant cross sections from the ground rovibrational level, while the transition to the
 $\text{A}\ ^1\Pi$ state makes very little contribution.
The behavior of the current cross sections are significantly different from those adopted in
\citet{heays17}.

Predissociation is possible following bound-bound transitions to the B $^1\Sigma^+$, 3 $^1\Pi$, and 4 $^1\Pi$ states.
Estimates of predissociation cross sections are computed for transitions to a wide range of bound rovibrational levels.
Cross sections for transitions from $v'',J'' = 0,0$ are shown in \autoref{fig:v0j0} computed using Equation~(\ref{pred}).
We find that the line cross sections due to predissociation are much smaller than the direct cross sections for the 4 $^1\Pi$ state.
However, while predissociation through the B $^1\Sigma^+$ and 3 $^1\Pi$ states give cross sections comparable to those of their direct continuum cross sections, the continuum cross section for the
2 $^1\Pi$ dominates the predissociation lines by more than
an order of magnitude over the relevant wavelength range. Predissociation does not appear to be important for the photodestruction of CS and is therefore not considered further.

\begin{figure}
    \centering
    \gridline{\fig{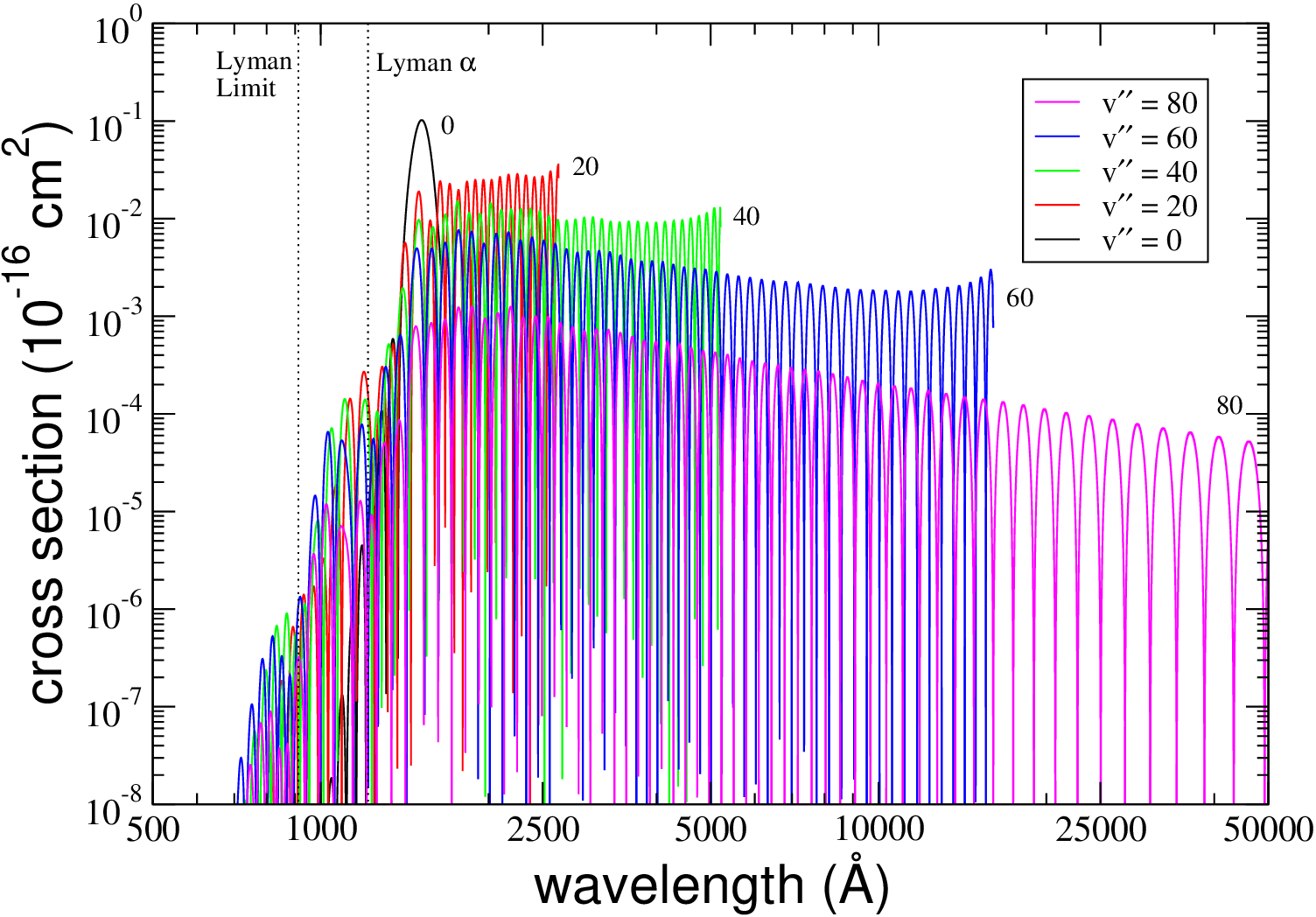}{0.5\textwidth}{(a)}
                  \fig{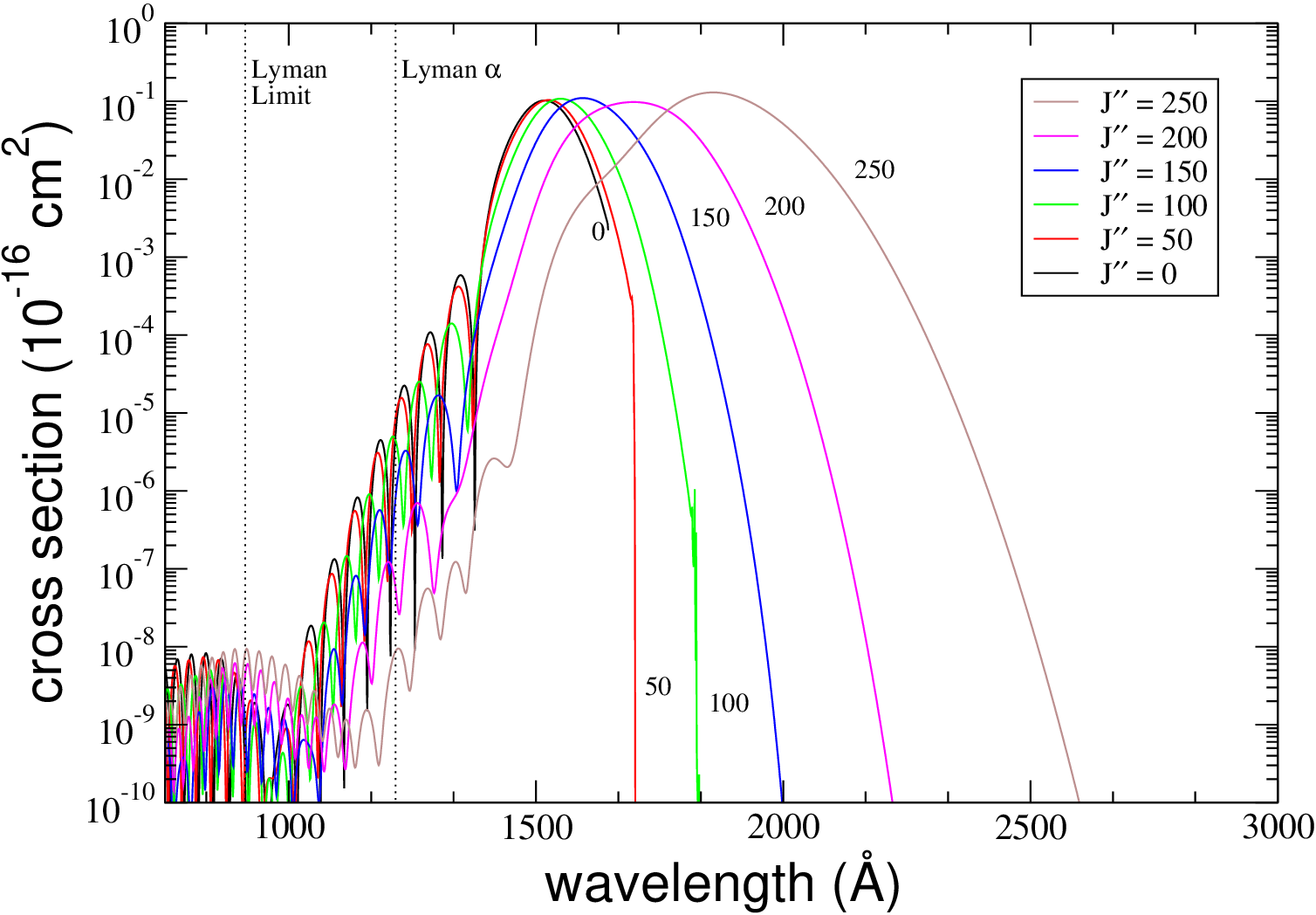}{0.5\textwidth}{(b)} }
    \caption{A sample of CS state-resolved cross sections for the 
		A$^\prime\ ^1\Sigma^+ \gets \text{X}\ ^1\Sigma^+$ 
		photodissociation transition. Transitions from initial 
		rovibrational levels (a) where $J'' = 0$ and 
		(b) where $v'' = 0$ are shown.}
    \label{fig:2Sigxc}
\end{figure}

The  $\text{A}^{\prime}\ ^1\Sigma^+ \leftarrow$ X$~^1\Sigma^+$ transition generally has large state-resolved 
cross sections; so a sampling of cross sections are displayed in \autoref{fig:2Sigxc}. Cross sections
 are plotted for several rotational levels of the ground vibrational level $v'' = 0$, and for several 
vibrational levels at their respective lowest rotational level, $J'' = 0$. State-resolved 
cross sections for the other five electronic transitions have also been computed (not shown).

\subsection{LTE Cross Sections}

\begin{figure}
    \centering
    \includegraphics[width=\textwidth]{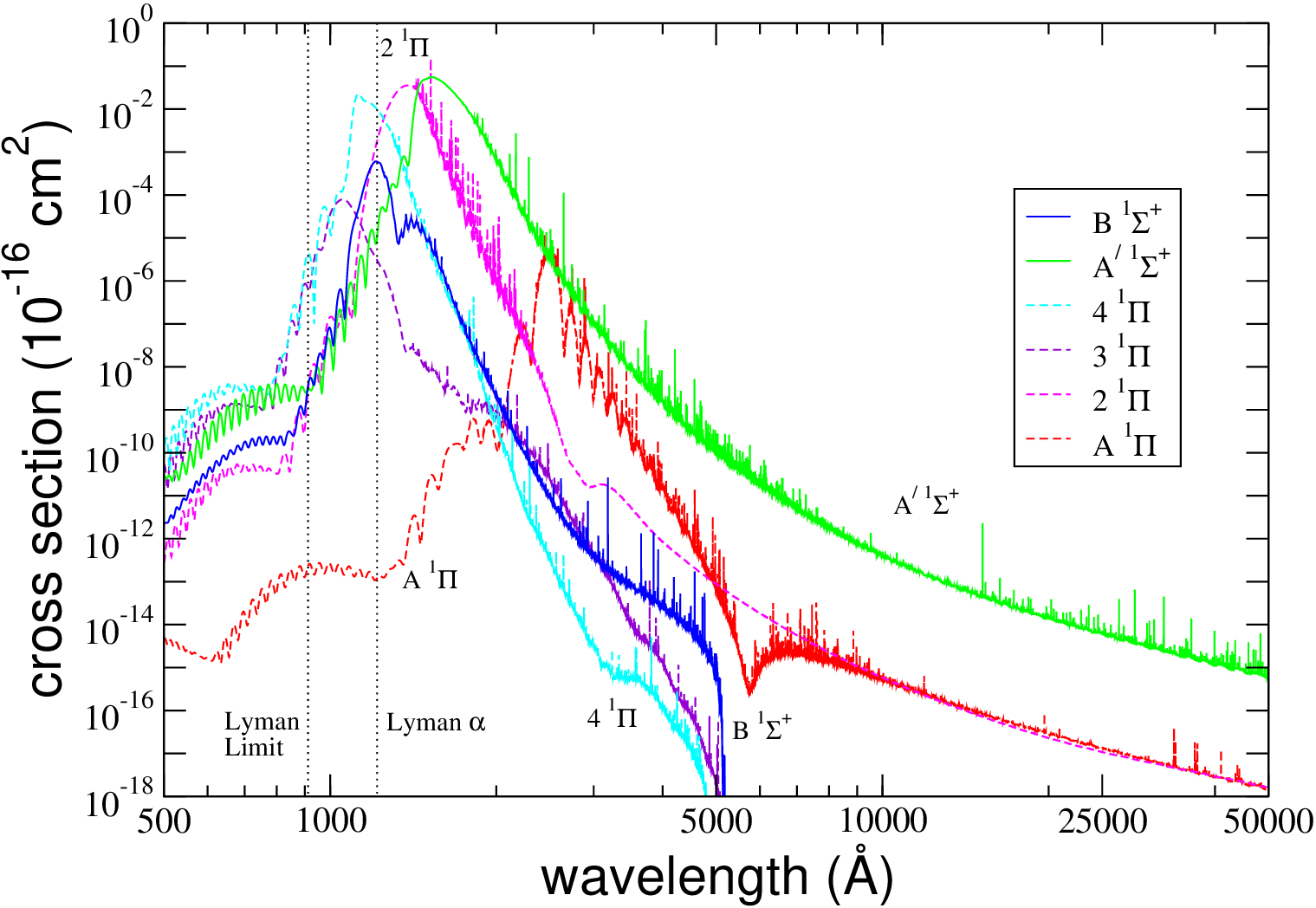}
    \caption{CS LTE cross sections at 3000 K for each of the six considered photodissociation transitions.}
    \label{fig:LTE3000K}
\end{figure}

\begin{figure}
    \centering
    \includegraphics[width=\textwidth]{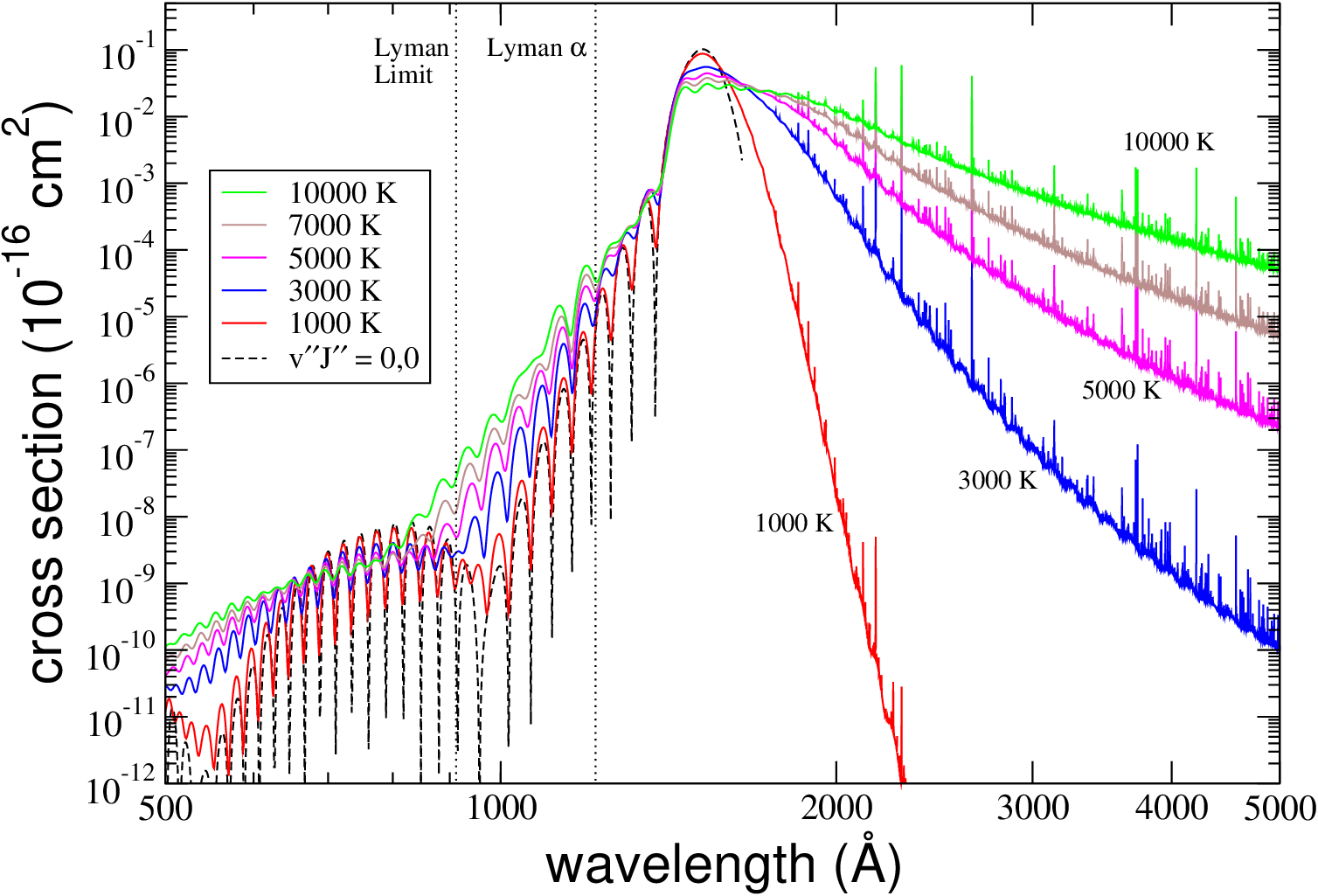}
    \caption{LTE cross sections for various kinetic temperatures for the $\text{A}^{\prime}\ ^1\Sigma^+ \gets \text{X}\ ^1\Sigma^+$
transition of CS. The $v'',J''=0,0$ state-resolved cross section is included as well for comparison.}
    \label{fig:2SigLTE}
\end{figure}

LTE cross sections have been computed for each transition using the state-resolved cross sections from 
1000 K to 10,000 K in 1000 K intervals. A comparison of LTE cross sections for each transition as a 
function of photon wavelength at 3000 K is displayed in \autoref{fig:LTE3000K}. 
The $\text{A}^{\prime}\ ^1\Sigma^+\leftarrow$ X$~^1\Sigma^+$ transition is the dominant transition at longer 
wavelengths, while the $4\ ^1\Pi\leftarrow$ X$~^1\Sigma^+$ transition dominates for short wavelengths. 
Since the $\text{A}^{\prime}\ ^1\Sigma^+ \leftarrow$ X$~^1\Sigma^+$ transition is dominant for the majority of 
wavelengths, LTE cross sections for this transition at several temperatures are shown in \autoref{fig:2SigLTE}.

\subsection{Photodissociation Rates}

Photodissociation rates for transitions from $v'',J''=0,0$ 
 for all six electronic transitions have been computed for the unattenuated 
ISRF and for the attenuated ISRF into interstellar clouds with total visual extinction. 
These values are listed and compared with those of \citet{heays17} and the UMIST
compilation \citep{umist} in Table~(\ref{photorate}). We consider fiducial diffuse and dense
clouds with total visual extinctions of $A_{\rm V}$ = 1 and 20, respectively. Consistent with the
cross section magnitudes, the ISRF photodissociation 
{ rates are}
dominated by the
A$^\prime~^1\Sigma^+ \leftarrow$ X$~^1\Sigma^+$ and 2 $^1\Pi \leftarrow$ X$~^1\Sigma^+$ transitions, which
leave the two atoms in their ground states. However,  about 10\% of the photodissociation yield
results in both C and S in their $^1D$ metastable states through the 4 $^1\Pi \leftarrow$ X$~^1\Sigma^+$ transition.
Using reliable CS photodissociation cross sections, the current unattenuated ISRF 
{ rates are}
about a factor
of $\sim$2.5-3 smaller than the estimates adopted by \citet{heays17} and \citet{umist}.

{
\floattable
\begin{deluxetable}{lccccccc}[ht!]
    \tablewidth{0pt}
    \tablecaption{Interstellar CS Photodissociation Rate Fits\tablenotemark{a}}.\label{table:photorate}
    \tablehead{Source & ISRF  &  Dense Cloud & & & Diffuse Cloud &  & Products\\
 \cline{3-4}  \cline{5-7}
     & $k$(s$^{-1}$) & $a_1$(s$^{-1}$) & $a_2$ &   $a_1$(s$^{-1}$) & $a_2$ & $a_3$ & C + S\\
     &  & [$a_4$(s$^{-1}$)] & [$a_5$] &    &  & & }
 \startdata
A $^1\Pi$ {$\leftarrow$} X & 1.50(-21) & 5.43(-22)  & 2.085 & 8.29(-22) & 3.73 & 4.00 & $^3P$ + $^3P$\\
2 $^1\Pi$ {$\leftarrow$} X & 1.35(-10) & 5.11(-11)  & 2.50 & 7.29(-11) & 4.16 & 4.37 & \\
A$^\prime~^1\Sigma^+$ {$\leftarrow$} X & 1.94(-10) & 7.59(-11)  & 2.19 & 1.08(-10) & 3.12 & 3.86 & \\
\hline
3 $^1\Pi$ {$\leftarrow$} X & 5.28(-15) & 1.86(-15)  & 3.16 & 2.75(-15) & 5.55 & 5.55 & $^1D$ + $^1D$ \\
B $^1\Sigma^+$ {$\leftarrow$} X & 9.56(-13) & 3.53(-13)  & 2.84 & 5.05(-13) & 4.89 & 5.00 & \\
4 $^1\Pi$ {$\leftarrow$} X & 4.05(-11) & 1.47(-11)  & 3.02 & 2.12(-11) & 5.25 & 5.30 & \\
\hline
Total & 3.70(-10) & 1.48(-10)  & 2.32 & 2.163(-10) & 3.98 & 4.21 & \nodata \\
& \nodata & [{2.13(-10)}]  & [{1.69}] & \nodata & \nodata & \nodata & \nodata \\
\hline
\hline
Heays\tablenotemark{b} & 9.49(-10) & 5.41(-10) & 2.49 & \nodata & \nodata & \nodata & \nodata \\
	     & \nodata & [9.49(-10)] & [1.95] & \nodata & \nodata & \nodata & \nodata \\
UMIST\tablenotemark{c} & 9.70(-10) & 9.70(-10) &  2.00 & \nodata & \nodata & \nodata & \nodata \\
\hline
\enddata
\tablenotetext{a}{{Fits to equations~\ref{fit1} and \ref{fit2}.}}
\tablenotetext{b}{\citet{heays17}.}
\tablenotetext{c}{\citet{umist}.}
\label{photorate}
\end{deluxetable}
}

\begin{figure}
    \centering
    \includegraphics[width=\textwidth]{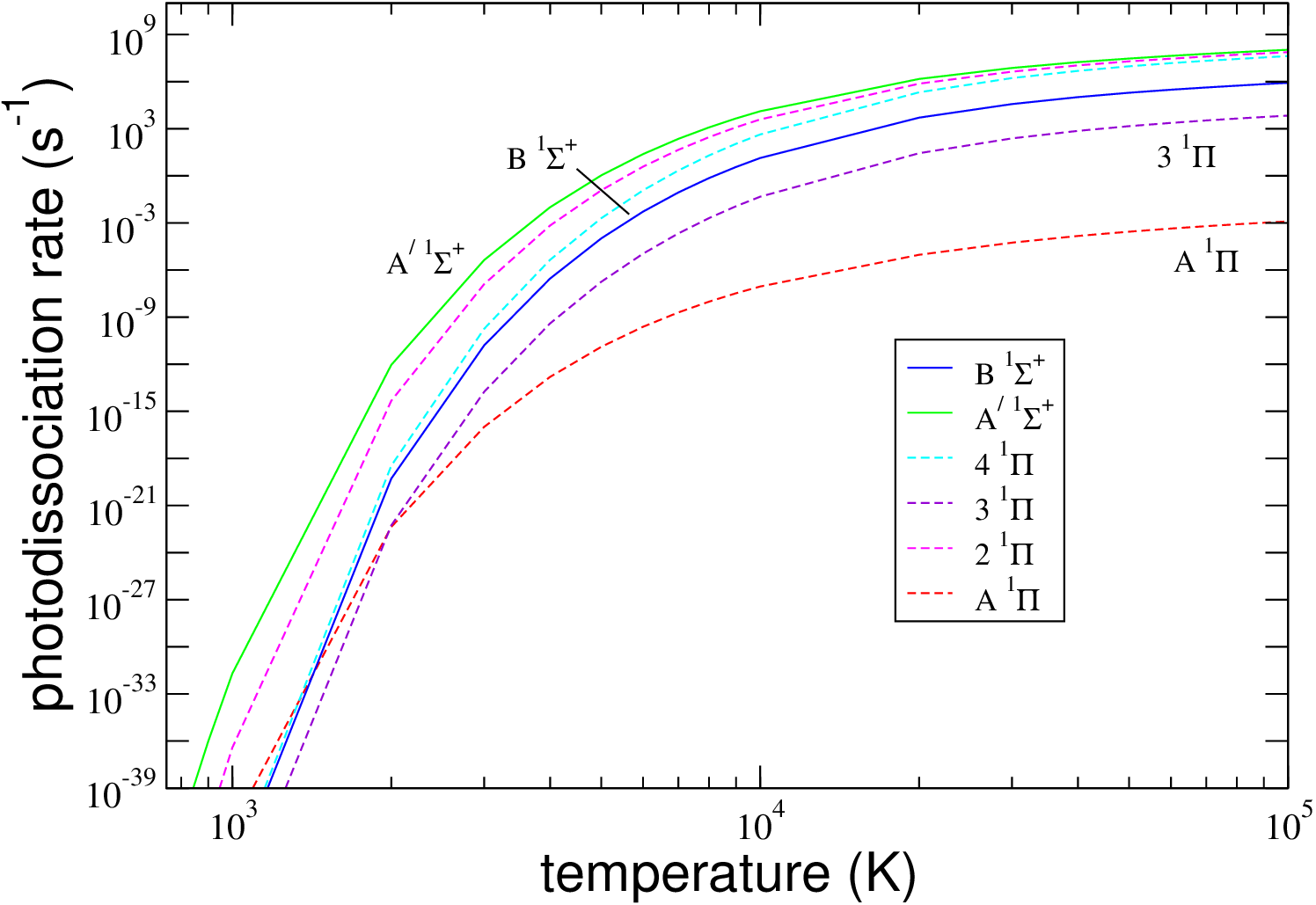}
  \caption{CS photorates in blackbody radiation fields for transitions from $v'',J''=0,0$ to each excited 
electronic state as a function of radiation temperature.}
    \label{fig:v0j0bbrate}
\end{figure}

{ We computed
photodissociation rates for a blackbody radiation field;} 
\autoref{fig:v0j0bbrate} 
shows a plot of the photodissociation 
{ rates}
when the molecule is initially in the $v'',J'' = 0,0$ rovibrational 
level for each final electronic state versus the blackbody temperature for a wide range of temperatures. 
Blackbody photodissociation rates were also obtained by \citet{heays17}, but they 
{ were}
normalized to reproduce the ISRF energy density from 912 to 2000 \AA, as opposed to the normalization
inherent in Eq.~(\ref{planckrad}) adopted here. Appropriate scale factors, e.g. geometric dilution, 
should be applied for the relevant astrophysical environment.
{ At the highest temperatures, the current photo rates should be taken as a lower limit as photoionization and photodissociation through high-lying Rydberg states will begin to become important.}

\begin{figure}
    \centering
    \includegraphics[width=\textwidth]{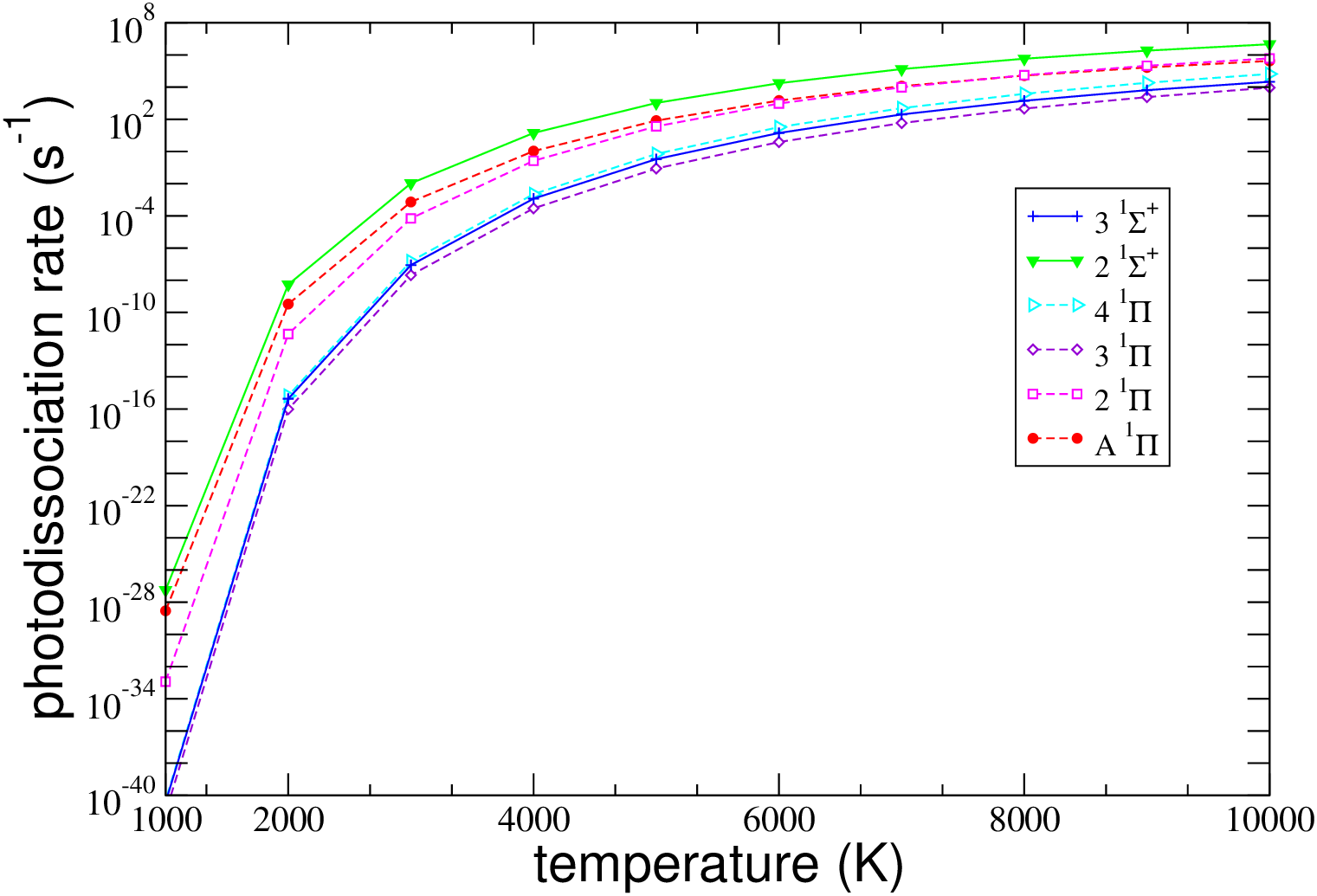}
  \caption{LTE blackbody photorates for each CS transition as
                   a function of temperature when the kinetic and
                   radiation temperatures are equal.}
    \label{fig:LTEbbrate}
\end{figure}

Finally, we consider a situation where a gas containing CS is in LTE at a certain temperature and is immersed 
in a radiation field generated by a blackbody at the same temperature (i.e., equal gas kinetic and
radiation temperatures). The photodissociation 
{ rates of CS in such a situation are}
computed using the LTE cross sections; a plot of these rates 
against the blackbody/gas temperature is shown in \autoref{fig:LTEbbrate}. 

\section{Conclusions}\label{sec:conclusions}

Accurate cross sections for the photodissociation of the CS molecule have been computed for transitions 
to several excited electronic states using new \textit{ab initio} potentials and transition dipole moment functions. 
The state-resolved cross sections have been computed for nearly all rotational transitions from vibrational 
levels $v''=0$ through $v''=84$ of the $\text{X}\ ^1\Sigma^+$ ground electronic state of CS. { Predissociation is found to be significantly smaller than direct photodissociation for CS.} Additionally,
 LTE cross sections have been computed for temperatures ranging from 1000 to 10,000 K. The computed 
cross sections are applicable to the photodissociation of CS in a variety of UV-irradiated interstellar 
environments including diffuse and translucent clouds, circumstellar disks, and protoplanetary disks. 
Photodissociation rates in the interstellar medium and in regions with a blackbody radiation field have 
been computed as well. To facilitate the calculation of local photorates for particular astrophysical 
environments, all photodissociation cross section data can be obtained from the UGA Molecular Opacity 
Project website.\footnote[5]{\url{http://www.physast.uga.edu/ugamop/}}

\acknowledgements
The work of R.J.P.\ and P.C.S.\ was supported by NASA grant NNX15AI61G. B.M.M.\ acknowledges support 
by the U.S.\ National Science Foundation through a grant to ITAMP at the Harvard-Smithsonian Center 
for Astrophysics under the visitor's program and Queen's University Belfast for a visiting research 
fellowship (VRF). The molecular structure calculations were performed at the National Energy Research 
Scientific Computing Center (NERSC) in Berkeley, CA, USA, and at the High Performance Computing 
Center Stuttgart (HLRS) of the University of Stuttgart, Stuttgart, Germany, where grants of time 
are gratefully acknowledged.  R.C.F.\ acknowledges support from NSF grant No.\ PHY-1503615.
{ ITAMP is supported in part by NSF grant No.\ PHY-1607396. We thank the referee for useful comments that improved the manuscript.}

\end{document}